\documentclass[prl,epsf,showpacs,twocolumn]{revtex4-1}
\usepackage[pdftex]{graphicx}
\usepackage{dcolumn}
\usepackage{bm}
\usepackage{epsfig}
\usepackage{latexsym}
\usepackage{amsmath}
\usepackage{amsfonts}
\usepackage{color}
\usepackage{array}
\usepackage{ulem}
\usepackage{braket}
\usepackage{dsfont}
\usepackage{mathrsfs}

\usepackage{hyperref}
\hypersetup{
    colorlinks,
    citecolor=blue,
    filecolor=red,
    linkcolor=magenta,
    urlcolor=red
}

\newcommand{\fref}[1]{Fig.\,\ref{#1}}
\newcommand{\eref}[1]{Eq.\,(\ref{#1})}

\newcommand{\beginsupplement}{%
        \setcounter{table}{0}
        \renewcommand{\thetable}{S\arabic{table}}%
        \setcounter{equation}{0}
        \renewcommand{\theequation}{S\arabic{equation}}%
        \setcounter{figure}{0}
        \renewcommand{\thefigure}{S\arabic{figure}}%
     }

\begin{document}


\title{A high fidelity light-shift gate for clock-state qubits}

\author{C. H. Baldwin}

\author{B. J. Bjork}
\email{bryce.bjork@honeywell.com}

\author{M. Foss-Feig}
\email{michael.feig@honeywell.com}

\author{J. P. Gaebler}

\author{D. Hayes}

\author{M. G. Kokish}

\author{C. Langer}

\author{J. A. Sedlacek}

\author{D. Stack}
\email{daniel.stack@honeywell.com}

\author{G. Vittorini}

\affiliation{Honeywell $|\!$ Quantum Solutions}

\begin{abstract}
To date, the highest fidelity quantum logic gates between two qubits have been achieved with variations on the geometric-phase gate in trapped ions, with the two leading variants being the M\o lmer-S\o rensen gate \cite{PhysRevLett.82.1971,PhysRevLett.117.060505} and the light-shift (LS) gate \cite{leibfried_2003,PhysRevLett.117.060504}.  Both of these approaches have their respective advantages and challenges.  For example, the latter is technically simpler and is natively insensitive to optical phases, but it has not been made to work directly on a clock-state qubit. We present a new technique for implementing the LS gate that combines the best features of these two approaches: By using a small ($\sim {\rm MHz}$) detuning from a narrow (dipole-forbidden) optical transition, we are able to operate an LS gate directly on hyperfine clock states, achieving gate fidelities of $99.74(4)\%$ using modest laser power at visible wavelengths.  Current gate infidelities appear to be dominated by technical noise, and theoretical modeling suggests a path towards gate fidelity above $99.99\%$.

\end{abstract}
\maketitle

The promise of quantum computers to efficiently solve certain classically intractable problems has spawned a broad experimental effort to realize scalable platforms for quantum computation.  At the very least, a good platform should offer high-fidelity state-initialization and readout, a high-fidelity universal gate set, a large ratio between the qubit coherence time and the gate times, and a reasonable path toward scaling to large numbers of qubits. Trapped ions appear very capable of meeting all of the above requirements, with state of the art experiments demonstrating combined state preparation and measurement fidelities above $99.9\%$ \cite{PhysRevLett.113.220501}, two-qubit gate fidelities of $99.9\%$ \cite{PhysRevLett.117.060505,PhysRevLett.117.060504}, and laser-based single-qubit gate fidelities above $99.99\%$ \cite{Baldwin_2019} (microwave-based single qubit gates have achieved even higher fidelities \cite{PhysRevLett.113.220501}).  Although these gate fidelities satisfy the fault-tolerance thresholds of some error correcting codes, the technical overhead to achieve them and the qubit overhead of performing error correction with them is still very high; scalable trapped-ion quantum computation likely requires further improvements of gate fidelity, reduced experimental overhead, or most likely both.

\begin{figure}[t]
\begin{center}
\includegraphics[width=0.98\columnwidth]{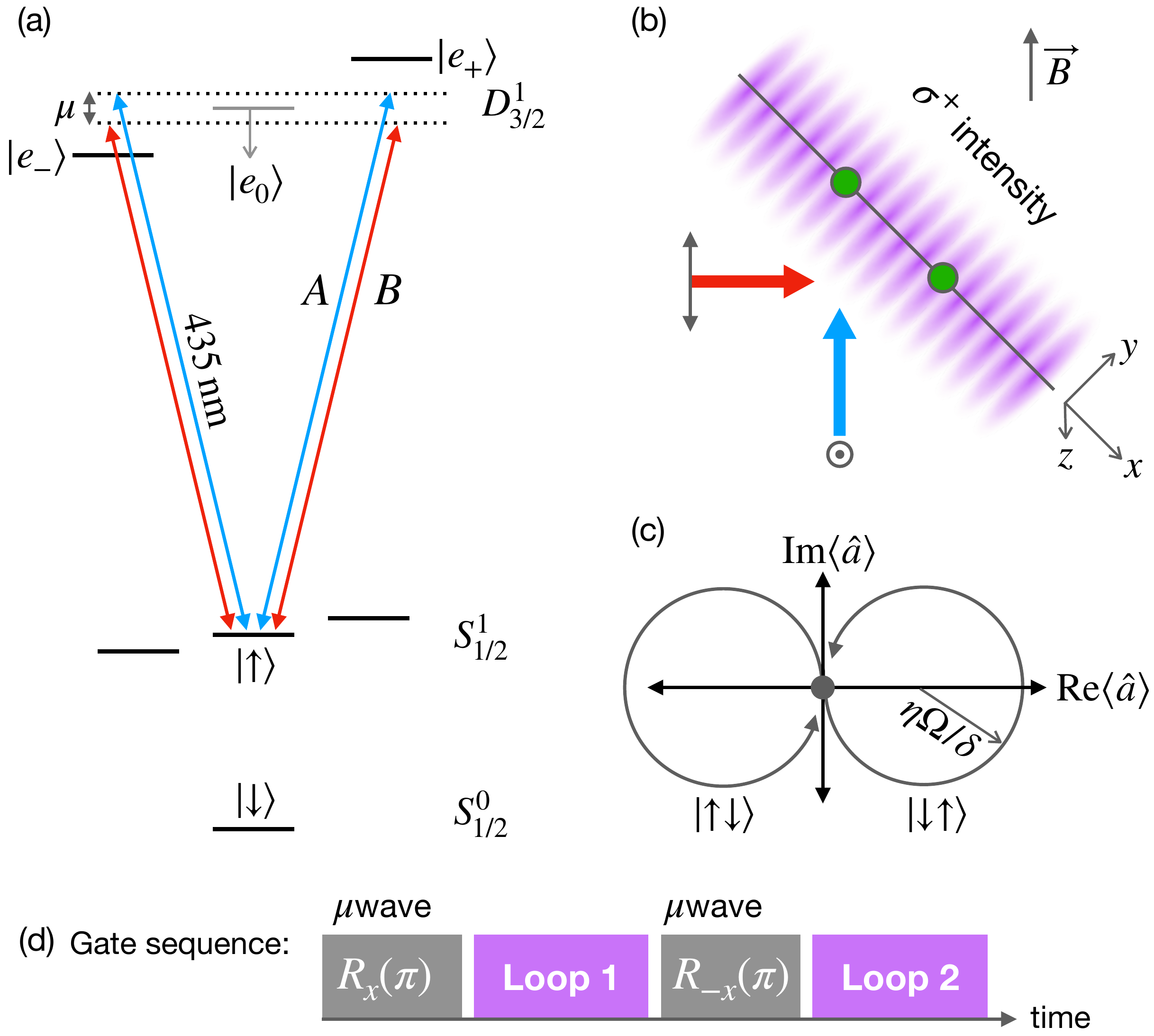}
\caption{(a) Level scheme used to implement a light-shift (LS) gate on a clock-state qubit. (b) Experimental setup: Due to the orthogonal polarizations there is no intensity modulation, but there is a polarization gradient that induces a spatial modulation of the AC-Stark shift experienced by an ion in the $\ket{\uparrow}$ state. (c) Gate operation: When the two ions are separated by an even integer multiple of $\pi / |\Delta\bm{k}|$ only the $\ket{\uparrow\downarrow}$ and $\ket{\downarrow\uparrow}$ states are driven, undergoing circular trajectories in phase space that return to their origin at a time $2\pi/\delta$. (d) We use a two-loop gate with spin echo in order to cancel any residual time-averaged AC-Stark shift during the gate.}
\label{fig:experimental_setup}
\end{center}
\end{figure}

The best two-qubit gate fidelities reported for trapped ions have been achieved using the geometric phase gate.  A spin-dependent force (SDF) is engineered, and drives excursions of a chosen collective mode of motion of a two-ion crystal.  This motion imparts spin-dependent geometric phases to the two-qubit wave function, and for suitably chosen parameters results in a maximally entangling gate $U_{zz}={\rm diag}(1,i,i,1)$.  There are several different microscopic realizations of the spin-dependent force underlying this gate, but the two-most successful approaches to date have been (1) The M\o lmer-S\o rensen (MS) gate, in which blue- and red-sideband transitions corresponding to a chosen collective mode of motion are driven simultaneously, and (2) The light-shift (LS) gate, in which a traveling wave generates a spatially and temporally modulated light shift of the qubit frequency (see \fref{fig:experimental_setup}).  Both gates can in principle be achieved on an optical qubit or between qubit states within the electronic ground-state manifold. The state-of-the-art fidelities reported in Refs.\,\cite{PhysRevLett.117.060505,PhysRevLett.117.060504} were both achieved in the latter context, generating couplings within the qubit manifold by coupling off-resonantly to a dipole-allowed $S\rightarrow P$ electronic transition.


The MS gate and LS gate both have benefits and draw backs.  For example, the former requires high-frequency ($\sim {\rm GHz}$) laser modulation, and one must be careful to remove the gate dependence on optical phases \cite{Lee_2005}, but it can be employed directly with qubits encoded in hyperfine clock states, which have very long coherence times.  The LS gate requires only low-frequency ($\sim {\rm MHz}$) laser modulation and is more natively insensitive to optical phases, but is incompatible with a clock-state qubit when operated off a dipole transition \cite{PhysRevLett.95.060502}.  Here, we demonstrate that a high-fidelity LS gate {\it can} in fact be operated on a clock-state qubit by utilizing a long-lived ($D$ rather than $P$) excited state. We build on an approach originally conceived in Ref.\,\cite{PhysRevA.76.040303}.  In that work the authors considered a scheme involving large ($\sim {\rm GHz}$) detunings from the $D$ state, which in turn required extremely high laser power.  Here we present a related gate scheme that works at much smaller  ($\sim {\rm MHz}$) detunings, and can be operated at essentially the same laser power used in more conventional approaches.  It combines many of the best features of the MS and LS gates while enabling a variety of technical simplifications, such as operating at much friendlier (visible rather than UV) wavelengths, for which power is more readily available, trap charging is mitigated \cite{doi:10.1063/1.3662118}, and integration of on-chip light delivery is easier \cite{10.1117/12.2288411}.

{\it Gate implementation.}---We implement the gate on two $^{171}{\rm Yb}^{+}$ ions, with qubit state $\ket{\downarrow(\uparrow)}$ encoded in hyperfine levels $\ket{F=1(0),m_F=0}$ of the $S_{1/2}$ manifold \footnote{A similar approach may work for other electronically similar ions (i.e. with a low-lying $D$ level).}. Two non-copropagating lasers with wave-vector difference $\Delta\bm{k}$ parallel to the trap axis and difference frequency $\mu$ are tuned $\pm\mu/2$ from resonance with the $\ket{\uparrow}(S_{1/2},\ket{F=1,m_F=0})\rightarrow \ket{e_0}(D_{3/2},\ket{1,0})$ transition (\fref{fig:experimental_setup}a).  The directions and polarizations of these two beams (\fref{fig:experimental_setup}b) are chosen such that $\Delta m_{F}=\pm1$ transitions are allowed while $\Delta m_F=0,\pm2$ are quadrupole-forbidden, thereby generating a coupling of $\ket{\uparrow}\rightarrow \ket{e_{\pm}}( D_{3/2},\ket{1,\pm1})$ while avoiding coupling to the $\ket{e_0}$ state \footnote{Note that the beam configuration chosen for this experiment and drawn in \fref{fig:experimental_setup}(b) is just one of many possibilities, e.g. the beams could be counter-propagating, and $\Delta \bm{k}$ could be oriented along a radial crystal mode.}.  Because the two gates lasers are detuned symmetrically between the $\ket{e_{\pm}}$ states, there is no time-averaged AC-Stark shift of the qubit-frequency, which greatly reduces the gate sensitivity to laser intensity fluctuations.  Despite the uniform intensity, a spatially modulated SDF is established by a polarization gradient, as shown in \fref{fig:experimental_setup}.  We choose the beat note $\mu$ to be near-detuned from the axial stretch mode ($\mu=\omega_{\rm str}+\delta$), in which the two ions move out of phase along the trap axis ($x$-direction in \fref{fig:experimental_setup}b), with creation(annihilation) operator $\hat{a}^{\dagger}(\hat{a})$.  Adiabatically eliminating the excited states, moving to the interaction picture with respect to the ion motion, dropping all terms rotating at frequencies faster than $\delta$, and making the Lamb-Dicke approximation, we obtain the following Hamiltonian describing an SDF applied to the two ions
\begin{align}
\label{eq:simple_hamiltonian} 
H\approx\eta\Omega\!\!\sum_{j=1,2}\!(-1)^j\big(1\!+\!\hat{\sigma}^z_j\big)\big(\hat{a}e^{i\delta t+i\varphi_j}+\hat{a}^{\dagger}e^{-i\delta t-i\varphi_j}\big).
\end{align}
Here $\Omega\approx g^2/\Delta$, with $g$ the magnitude of the single-photon Rabi frequency for the $\ket{\uparrow}\rightarrow\ket{e_{\pm}}$ transitions (chosen to be equal for both lasers) and $\Delta/(2\pi)\approx B\times1.4\,{\rm MHz}/{\rm Gauss}$ the detuning of these levels from the average frequency of the lasers, and $\eta = |\Delta\bm{k}|\sqrt{\hbar/(4 m\omega_{\rm str})}$ is the Lamb-Dicke parameter for the gate mode (with $m$ the ion mass). The ion dependent motional phases are $\varphi_j=\varphi+|\Delta\bm{k}|x_{j}$, with $\varphi$ the optical (Raman) phase and $x_j$ the mean position of ion $j$ along the trap axis, and the factor of $(-1)^j$ reflects the antisymmetry of the gate mode. If the ions are spaced by an even integer multiple of $\pi/|\Delta\bm{k}|$ then the SDF is in phase on the two ions.  The (antisymmetric) gate mode is therefore not driven when the ions are in the $\ket{\uparrow\uparrow}$ state. The $\ket{\downarrow\downarrow}$ state is trivially decoupled by being far from resonance, so only the $\ket{\uparrow\downarrow}$ and $\ket{\downarrow\uparrow}$ states are driven in this configuration, as shown in \fref{fig:experimental_setup}c. At a time $2\pi/\delta$ the spin and motion decouple, imprinting the unitary $U(\Phi)=R_z(\epsilon){\rm diag}(1,e^{i\Phi},e^{i\Phi},1)$ on the spin wavefunction, where $\Phi=2\pi(\Omega\eta/\delta)^2$.  Here $R_z(\epsilon)$ is a small unwanted spin rotation about the $z$ axis due to either AC-Stark shifts from the $D_{3/2}^2$ manifold or fluctuations of the mean laser frequency.  We implement a two-loop gate with a spin-echo to remove the effect of this small residual shift.  Adjusting the laser power such that $\Phi=\pi/4$ then yields the maximally entangling gate $R_{x}(\pi)U(\pi/4)R_{-x}(\pi)U(\pi/4)={\rm diag}(1,i,i,1)$.  It is important to note that several approximations leading to \eref{eq:simple_hamiltonian} are not that well justified.  In particular, perturbative adiabatic elimination of the excited states is a fairly crude approximation, with excited state populations on the order of several percent being typical for the experimental parameters used here.  In the supplemental material we give the Hamiltonian with the excited states included, which is used for all numerical simulations.  Importantly, we show that shaping the SDF pulses can reduce residual excited state population at the end of the gate to significantly below the $10^{-4}$ level.

To understand why the gate can be operated with high fidelity at modest laser powers, it is helpful to first consider the effect of the excited state linewidth on laser-based trapped ion gates in general (these considerations apply to both the MS gate and the LS gate).  Gate fidelities are limited in part by technical issues, such as fluctuations of experimental parameters, which in general are more detrimental for longer gates, and by off-resonant couplings that are more detrimental for short gates.  It therefore makes sense to demand some fixed gate time $t_{\rm g}$ that minimizes these errors, which fixes the size of $\delta=2\pi/t_{\rm g}\sqrt{K}$ (for a gate comprised of $K$ successive phase space loops); optimal values of $t_{\rm g}$ are typically of order $10$'s of $\mu$s.  Since $\delta$ must be proportional to the SDF in order to achieve the desired phase $\Phi$, we have $g^2/\Delta\propto t_{\rm g}^{-1}$ \footnote{This entire discussion holds only for $g\lesssim\Delta$, which is the regime we are interested in.  We note that a similar gate can be operated even outside this regime, as shown in Ref.\ \cite{PhysRevA.77.050303}.}. Remembering from Fermi's golden rule that $g^2\propto P\Gamma$, with $P$ the laser power and $\Gamma$ the excited state linewidth, this implies a proportionality constraint $P(\Gamma/\Delta)\propto t_{\rm g}^{-1}$.  A more fundamental source of gate error for laser-based gates is spontaneous emission from the excited state: For an excited state decay rate $\Gamma$, the gate error incurred due to spontaneous emission obeys $\varepsilon\propto \Gamma(g^2/\Delta^2)t_{\rm g}\propto(\Gamma/\Delta)$.  Thus demanding a fixed minimum infidelity of $\varepsilon$ fixes the ratio $\Gamma/\Delta$.  Substituting this back into the gate-time constraint we find that the laser power should satisfy \footnote{The required power also has a wavelength dependence that has been ignored in this expression, and slightly favors the longer wavelengths encountered in a $D$-state gate relative to a $P$-state gate.}
\begin{align}
\label{eq:power}
P\propto 1/(\varepsilon t_{\rm g}),
\end{align}
which is {\it independent} of the linewidth $\Gamma$.  There is one catch: since $g^2/\Delta$ is fixed by the gate time, the excited state population obeys $P_e\propto (g/\Delta)^2\propto1/(t_{\rm g}\Delta)$.  As the linewidth narrows, $\Delta$ must decrease in proportion to maintain the gate time at fixed laser power, implying that the excited state population must grow.  This behavior is the underlying reason why adiabatic elimination of the excited state is not a particularly good approximation given modest laser power, as mentioned earlier.

\begin{figure}[t]
\begin{center}
\includegraphics[width=0.7\columnwidth]{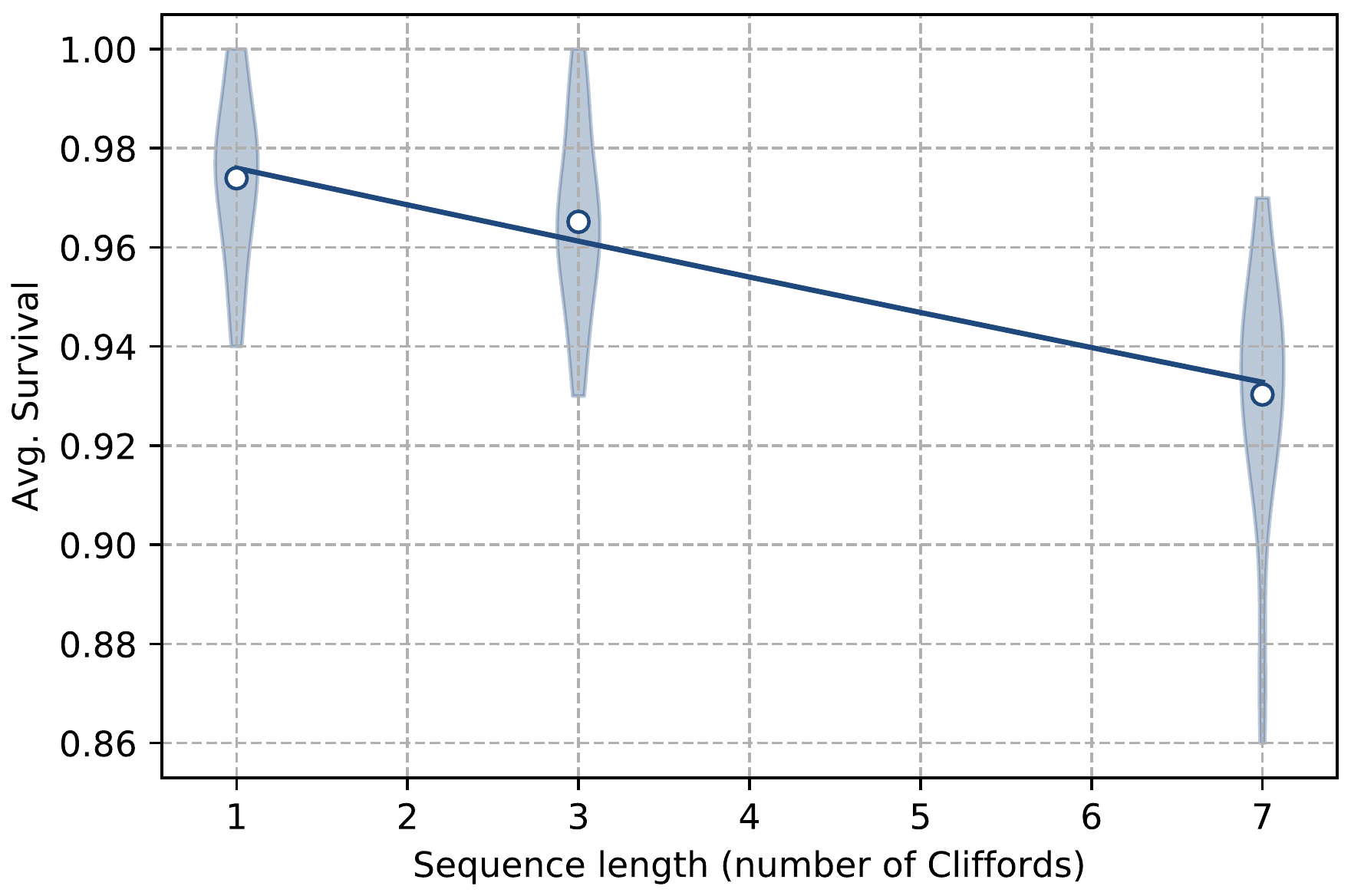}
\caption{Experimental gate fidelity measurement using the procedure described as SRB-{\it lite} in Ref.\,\cite{Baldwin_2019}.  The width of the shaded regions reflects the fraction of random Clifford sequences resulting in the corresponding average survival, and the line is from an exponential fit (see supplemental material for details).  Making conservative assumptions described in the supplemental material, we estimate the average gate fidelity to be $99.74(4)\%$.}
\label{fig:gate_results}
\end{center}
\end{figure}

{\it Experimental gate demonstration.}---Two $^{171}$Yb$^{+}$ ions are loaded into a nearly-harmonic trap with frequencies $(\omega_{x},\omega_y,\omega_z)=2\pi\times(1.16,2.57,3.05)\,{\rm MHz}$ by Doppler cooling in a $5.57$\,Gauss magnetic field.  The qubit states $\ket{\uparrow(\downarrow)}$ are encoded in the $^2 S_{1/2}$ hyperfine levels $\ket{F=1(0),m_F=0}$ of the $^{171}$Yb$^+$ electronic ground state, where $F$ and $m_F$ are the total angular momentum and its projection along the quantization axis (see \fref{fig:experimental_setup}a). The Yb ions are initialized to $\ket{\downarrow}$ via optical pumping \cite{PhysRevA.76.052314}. We then Raman sideband cool the axial stretch mode (which serves as our gate mode) to a motional occupation of $\bar{n}\sim$ 0.1 quanta, and all other modes to $\bar{n}<0.2$.  Global single qubit operations are realized via microwave radiation resonant with the hyperfine frequency of approximately $12.6428\,$GHz, and an average single-qubit gate error of $9(6)\times 10^{-5}$ for a single $^{171}$Yb$^+$ ion is measured by randomized benchmarking \cite{Hayes_2019}.  Two-qubit operations are realized via two beams propagating at a relative angle of $90^{\circ}$, such that the wave vector difference $\Delta \bm{k}$ is parallel to the axis of the two-ion crystal, and detuned from each other by $\mu =  \omega_{\rm str} + \delta$.  To calibrate the gate, we first measure the axial trap frequency and then set $\mu$ based on the desired gate time. We use a two-loop gate with spin echo (\fref{fig:gate_results}b) for a total gate time of $t_{\rm gate}=2(t_{\rm loop}+t_{\pi})$, and choose $t_{\rm loop}=45\,\mu$s, which in turn fixes $\delta=2\pi/t_{\rm loop}\approx 2\pi\times 22.2\,$kHz.  The gate fidelity is then optimized by tuning the laser power, which in general is about $50\,$mW per beam for a spot size of $\approx35\,\mu$m at the ion positions. For each loop, the laser intensity is ramped on and off with a $\sin^2(t)$ profile over a $2\,\mu$s duration \footnote{The loop time is defined as the time between the middle of the pulse rise and the middle of the pulse fall.  While $\delta = 2\pi/ t_{\rm loop}$ is not exactly optimal for non-zero rise time, it is an excellent approximation and contributes no appreciable error for a two-loop gate.}. After gate operations, the ions are illuminated with a laser beam resonant with the $5^2 S_{1/2} \rightarrow 5^2P_{1/2}$ transition, and the qubit state is inferred from the resulting fluorescence \cite{PhysRevA.76.052314}.  For a detection duration of $400\,\mu$s, we detect on average approximately 9 photons for each ion in $\ket{\uparrow}$ and 0.1 photons for each ion in $\ket{\downarrow}$, leading to a detection fidelity of $\sim99\%$.   
\

In the apparatus used for this work, we do not have easy access to high-resolution individual addressing, making full randomized benchmarking challenging.  However, arbitrary {\it global} rotations can be performed using microwaves, and together with the LS gate give us access to the complete symmetric subspace of the two-qubit Hilbert space spanned by $\{\ket{\downarrow\downarrow},(\ket{\downarrow\uparrow}+\ket{\uparrow\downarrow})/\sqrt{2},\ket{\uparrow\uparrow}\}$.  We therefore benchmark the gate fidelity experimentally using the subspace randomized benchmarking protocol of Ref.\ \cite{Baldwin_2019} (specifically the protocol described therein as SRB-\textit{lite}) restricted to this symmetric subspace.  The procedure consists of applying random sequences of qutrit-Clifford gates with varying sequence lengths. Each sequence is designed to return the qubits to a known state in the absence of errors, and from the decay of the return probability with sequence length one can extract the process fidelity across the symmetric subspace of the gates involved.  The error of the LS-gate can be extracted from the Clifford-gate fidelity (under certain reasonable assumptions \cite{Baldwin_2019}) by simple counting arguments.  From the data in \fref{fig:gate_results}b, we estimate an LS-gate average fidelity of $99.74(4)\%$.


\

\

{\it Theoretical fidelity limit and experimental errors.}---Non-technical errors associated with this gate are mostly similar to those encountered in operating either an LS or MS gate on a $P$-state transition, namely: (1) deviations from the Lamb-Dicke regime, (2) spontaneous emission, and (3) failure of the various rotating-wave approximations (i.e. the presence of off-resonant couplings).  Errors due to (1) are essentially the same as for $P$-state gates, and can be suppressed to below $10^{-5}$.  Errors due to (2) and (3) require additional scrutiny.  In everything that follows, we define the gate error as $\epsilon=1-\mathscr{F}_{\rm avg}$, with $\mathscr{F}_{\rm avg}$ being the average fidelity.

Raman scattering is an important error mechanism for all optical gates, and while it can in principle be made extremely small given enough laser power, in practice it is often quite important (contributing nearly half of the $\approx 10^{-3}$ infidelity in Ref.\,\cite{PhysRevLett.117.060505}).  Because the $D$-state gate must operate at single photon detunings $\Delta\ll\Delta_{\rm hf}$ to achieve reasonable gate speeds at feasible laser powers, any photon scattering event (Raman or Rayleigh) off the $D$-state unambiguously determines the qubit to be in the $\ket{\uparrow}$ state, so we take the associated error to be simply the total scattering rate multiplied by the gate time.  This approach gives $\epsilon_{\gamma}^{D}\approx 2\pi(\Gamma/\Delta)\times\frac{\sqrt{K}}{\eta}$.  For $100\,$mW total laser power focused to a $35\,\mu {\rm m}$ spot size, we find that a  $100\,\mu{\rm s}$ $2$-loop gate incurs an error $\epsilon^{D}_{\gamma}\approx 6\times 10^{-5}$.  As for any light shift gate, this error can in principle be made smaller by using more laser power and detuning further. Ultimately the spontaneous emission limit is set by Raman scattering off the $P$ states, which causes an error $\epsilon_{\gamma}^{P}$ that increases for larger $\Delta$ assuming the gate time is held constant. Optimizing over $\Delta$, we find that a minimum spontaneous emission error $\epsilon_{\gamma}=\epsilon^D_{\gamma}+\epsilon^P_{\gamma}\approx 7 \times10^{-6}$ is achievable in principle, though we note that nearly $2\,$W of laser power would be required for the same laser spot size.

\begin{figure}[t]
\begin{center}
\includegraphics[width=0.9\columnwidth]{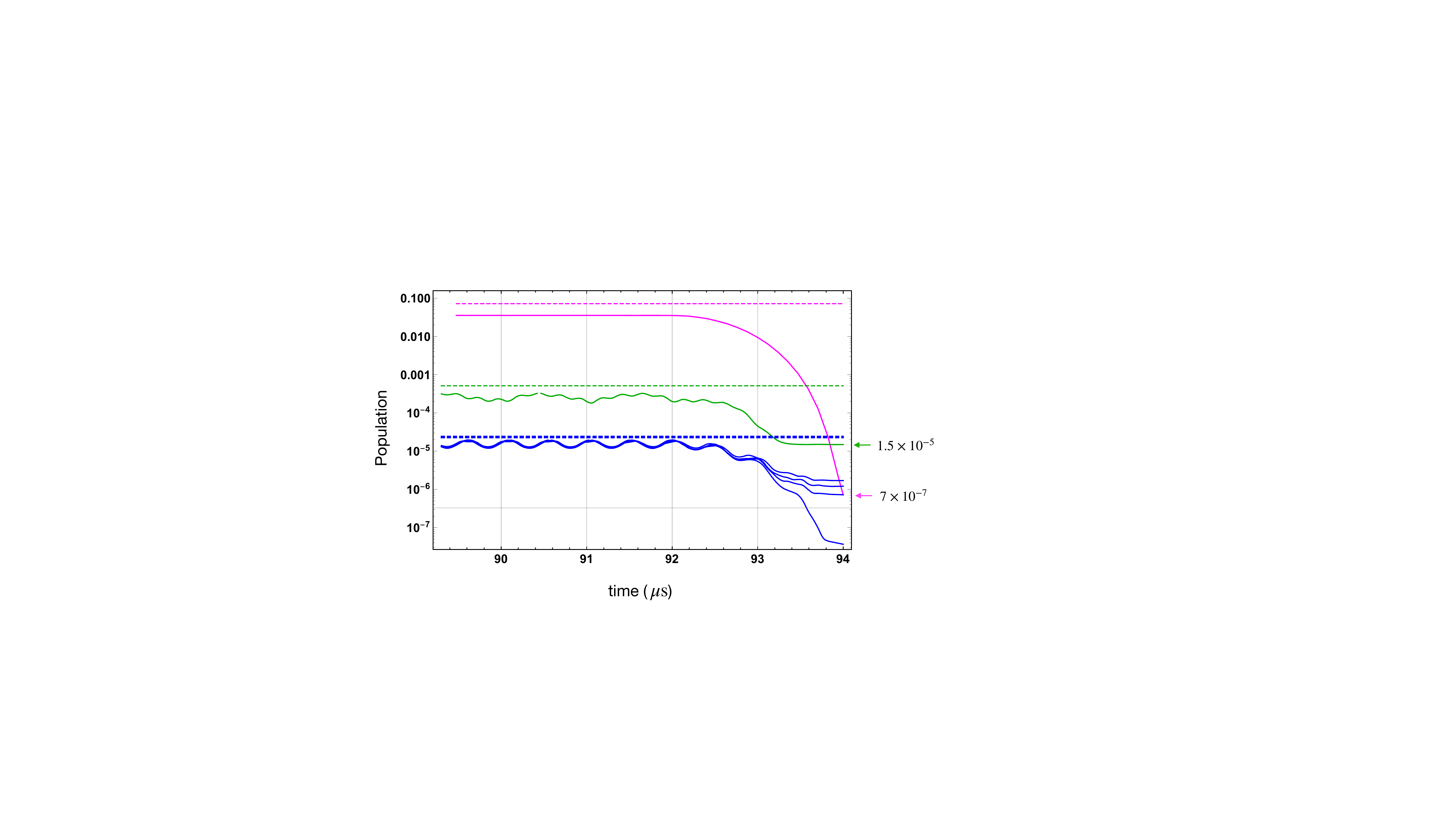}
\caption{Contributions to the gate infidelity from off-resonant couplings to: radial modes (blue), the axial c.o.m.\ mode (green), and the $D$-states (purple).  All of these curves have been computed numerically, and filtered with a moving-average to remove structure at the time scale of $1/\Delta$ (for visual clarity).  The gate starts at time $t=0$, but we only show the final $\sim 5\,\mu{\rm s}$ in order to resolve the effect of pulse shaping in reducing transient populations. Dashed lines are theoretical estimates in the absence of pulse shaping; with pulse shaping (solid lines) the total errors from all such couplings can be brought well below $10^{-4}$, with the single largest contributor being the c.o.m. coupling.}
\label{fig:off_resonant_pops}
\end{center}
\end{figure}

There are two types of off-resonant couplings that limit the gate fidelity in this scheme: two-photon couplings that oscillate near motional frequencies, and single-photon couplings that oscillate at or near the single-photon detuning $\Delta$.  The former are dominated by a time-dependent light shift (with no coupling to the motional modes), with the next-leading contribution being coupling to the axial center of mass (c.o.m.) mode.  The latter are dominated by carrier transitions directly to the $\ket{e_{\pm}}$ states, but there are also non-negligible contributions from coupling to the blue sidebands of $\ket{e_{-}}$ \footnote{Near-ground-state cooling makes coupling to red sidebands of $\ket{e_{+}}$ much less important.  Note that heating of spectator modes that occurs during longer gate sequences in SRB may lead to enhanced off-resonant coupling errors, which increase with the mean occupation number of those modes}.  For any of these couplings, we can estimate the induced gate error as simply being equal to the population transfer into the associated state.  For a square gate pulse, these populations can in turn be estimated analytically using second-order perturbation theory, and are shown as dashed lines in \fref{fig:off_resonant_pops}.  While some of these errors can be quite large, they can be significantly suppressed by shaping the gate pulses.  The solid lines in \fref{fig:off_resonant_pops} show the associated populations during the gate, along with their final values, when the gate beams are turned on and off using a $\sin^2(t)$ profile with a duration of $2\,\mu s$, as in the experiment.  The theoretical gate error estimates reported in the top of Tab.\,\ref{tab:errors} are obtained by numerical simulation of the gate Hamiltonian without making the Lamb-Dicke approximation, assuming ground-state cooling of all modes, and including all of the off-resonant couplings described above; at our current laser power and beam waists, we estimate that $\epsilon<10^{-4}$ should be possible, with significant further improvements possible with more laser power.

\begin{table}[!h]
  \begin{center}

    \begin{tabular}{l|c} 
      \textbf{Mechanism} & \textbf{Gate error ($\times 10^{-4}$)} \\
       \hline
       \hline
      Spontaneous emission & $0.6$ \\
      Off-resonant + L.D. errors & $ 0.2$\\
      \textbf{Total} & $0.8$ \\
      \textbf{Min} & $0.27$ \\
       \hline
      Leakage &  $\sim 3$  \\
      Laser phase noise & $\sim 3$ \\
      Gate mode heating $~~$ & $<4$ \\
      c.o.m. heating &$$1\\
      Microwaves & $2$ \\
      \textbf{Total} & $\epsilon\lesssim 12$ \\
    \end{tabular}
   \caption{ \label{tab:errors}Top section: Non-technical contributions to the total gate error (see text for definitions) for the parameters and pulse-shaping used in the experimental gate demonstration. \textbf{Total} indicates the expected gate infidelity due to non-technical errors for the experimentally available laser power ($100\,{\rm mW}$), while \textbf{Min} refers to the theoretical lower limit limit, which requires $2\,{\rm W}$ of laser power. Bottom section: estimated contribution of various sources of technical noise to the total experimentally measured gate infidelity.
   }
  \end{center}
\end{table}

The experimental gate error of $2.6(4)\times 10^{-3}$ is considerably larger than the theoretical limit.  In the bottom section of Tab.\,\ref{tab:errors} we list technical sources of error that we are aware of, which account for approximately half of the experimentally measured gate error (we believe that SRB-{\it lite} likely underestimates the actual gate fidelity, as the return probability in \fref{fig:gate_results} gets pulled down by heating during long gate sequences, causing the fit to overestimate the initial per-gate error). Contributions to the error described as ``leakage'' come from both spontaneous emission {\it and} any residual population remaining in the $D$ states at the end of the gate; causes of this residual population are due in large part to a small coupling of $\ket{\uparrow}\rightarrow\ket{e_0}$ that exists for imperfect polarization of the beam pointing along the quantization axis.  Laser phase noise contributes to the gate error primarily by inducing fluctuations in the time-averaged AC-Stark shift during the gate, which the spin-echo fails to perfectly null.  A more detailed discussion of these error sources and estimates can be found in the supplemental material.

{\it Outlook.}---The relatively low technical overhead in implementing a $D$-state LS gate provides a promising path towards high-fidelity gates in a scalable architecture.  In particular, we believe that modest improvements to both trap noise and laser line width should enable two-qubit gates operating at fidelities approaching or even exceeding $99.99\%$ fidelity.  Ideally, one would like to implement a universal gate set while maintaining the technical simplifications of this two-qubit gate (i.e. without introducing UV lasers or GHz frequency modulation).  In principle, one could combine global microwave rotations with local $z$-rotations induced by the gate lasers to implement arbitrary single-qubit rotations, though understanding the feasibility and achievable fidelity with this approach would need further study.  Alternatively, at the added cost of modulating the gate lasers at the hyperfine splitting one could readily implement high-fidelity all optical single-qubit gates.

{\it Acknowledgements.}---We thank Patty Lee and Hartmut H\"affner for helpful discussions.

\

\

\

\

\

\

\

\

\

\

\appendix

\section{SUPPLEMENTAL MATERIAL}
\beginsupplement

\subsection{Approximations leading to the gate Hamiltonian}

In the manuscript, for simplicity we wrote the gate Hamiltonian after a number of approximations had been made, some of which are not necessarily that well justified.  Here we give the microscopic Hamiltonian used for our numerical simulations, and describe the approximations resulting in \eref{eq:simple_hamiltonian} in more detail.  Ignoring terms that rotate at optical and hyperfine (GHz) frequencies, but keeping all optical excited states and motional modes, the Hamiltonian can be written $H=H_{\rm phonon}+H_{\rm ion}+H_{\rm laser}$, where
\begin{widetext}
\begin{align}
H_{\rm phonon}&=\sum_{\alpha,\nu}\omega_{\alpha,\nu}\hat{a}^{\dagger}_{\alpha,\nu}\hat{a}^{\phantom\dagger}_{\alpha,\nu}\label{eq:fullham_init}\\
H_{\rm ion}&=\Delta \sum_{j=1,2}\big(\ket{e_{+,j}}\bra{e_{+,j}}-\ket{e_{-,j}}\bra{e_{-,j}}\big)\\
H_{\rm laser}&=\sum_{j=1,2}\sum_{\tau=\pm}\bigg(\big(g_{A}^{\tau}e^{-i(\bm{k}_{A}\cdot \hat{\bm{r}}_j+\frac{\mu}{2}t+\phi_{A})}+g_{B}^{\tau}e^{-i(\bm{k}_{B}\cdot \hat{\bm{r}}_j-\frac{\mu}{2}t+\phi_{B})}\big)\ket{e_{j,\tau}}\bra{\uparrow_j}+{\rm h.c.}\bigg).\label{eq:fullham_fin}
\end{align}
\end{widetext}
Here $\omega_{\alpha,\nu}$ is the frequency of the $\alpha^{\rm th}$ normal mode of motion of the two ion crystal along the $\nu^{\rm th}$ principal axis of the trapping potential, and $\hat{a}^{\phantom\dagger}_{\alpha,\nu}(\hat{a}^{\dagger}_{\alpha,\nu})$ is the annihilation(creation) operator for that mode.  The optical phases of lasers $A$ and $B$ are denoted  $\phi_A$ and $\phi_B$ (see \fref{fig:experimental_setup} of the manuscript for a definition of the two beams), and we have chosen to work in a frame for which $\ket{\uparrow}$ and $\ket{e_0}$ have zero energy.  The matrix element $g^{\tau}_{A(B)}$ is proportional to the electric field amplitude of laser $A(B)$, and (because this is a quadrupole transition) depends on both the polarization {\it and} wave-vector of that laser.  Note that we do ignore coupling to the $\ket{e_{0}}$ state in \eref{eq:fullham_fin}, and also to the $F=2$ states of $D_{3/2}$.  Imperfections of the former approximation are discussed in the section titled ``Estimates of experimental error''.  The latter generates an AC-Stark shift of the qubit frequency on the order of $1\,{\rm kHz}$; for a two loop gate with spin echo, and considering the intensity stability of our lasers, we find that this shift should not contribute an appreciable error.

It is common to perturbatively eliminate the excited states, under the assumption that $g\ll\Delta$.  Assuming $|g_{A}^{\tau}|=|g_{B}^{\sigma}|\equiv g$ and using the fact that, for our chosen beam directions and polarizations, $\arg(\{g_{A}^{-},g_{A}^{+},g_{B}^{-},g_{B}^{+}\})=\{0,0,-\pi/2,\pi/2\}$, this procedure yields $H\approx H_{\rm phonon}+H_{LS}$, with
\begin{align}
H_{LS}=\frac{4g^2\Delta}{\Delta^2-\mu^2/4}\sum_{j=1,2}\ket{\uparrow_j}\bra{\uparrow_j}\sin\big(\Delta\bm{k}\cdot\hat{\bm{r}}_j+\mu t+\delta\phi\big).
\label{eq:full_ls_ham}
\end{align}
Here $\delta\phi\equiv\phi_{A}-\phi_B$ is the phase of the beat note between the two gate lasers.  To arrive at \eref{eq:simple_hamiltonian} in the main text, we rewrite the $\nu^{\rm th}$ cartesian component of the position operators (decomposed in a basis determined by the crystal principal axes) in terms of creation/annihilation operators for the crystal normal modes along that direction as
\begin{align}
\hat{r}^{\nu}_j = \langle \hat{r}^{\nu}_j\rangle+\sum_{\alpha}\eta_{j}^{\alpha,\nu}(\hat{a}^{\dagger}_{\alpha,\nu}+\hat{a}^{\phantom\dagger}_{\alpha,\nu}).
\end{align}
Here $\eta_{j}^{\alpha,\nu}=\Delta k^{\nu}B_{j}^{\alpha,\nu}\sqrt{\hbar/(2m\omega_{\alpha,\nu})}$, $B_{j}^{\alpha,\nu}$ is the component of the $\alpha^{\rm th}$ normal mode along the $\nu^{\rm th}$ principal axis onto the $j^{\rm th}$ ion, and $\omega_{\alpha,\nu}$ is the frequency of that mode.  Now expanding the $\sin()$ function in \eref{eq:full_ls_ham} to lowest order in the Lamb-Dicke parameters, using $\Delta/(\Delta^2-\mu^2/4)\approx 1/\Delta$, and defining $\Omega=g^2/\Delta$, we find
\begin{align}
H_{\rm LS}&\approx 2\Omega\sum_{j=1,2}(1+\hat{\sigma}^z_j)\sin(\mu t+\varphi_j)\\
&+2\Omega\sum_{j=1,2}(1+\hat{\sigma}_j^z)\cos(\mu t+\varphi_j)\sum_{\alpha,\nu}\eta_{j,\nu}(\hat{a}^{\phantom\dagger}_{\alpha,\nu}+\hat{a}^{\dagger}_{\alpha,\nu}).\nonumber
\end{align}
The first term is a time-dependent AC-Stark shift that nominally contributes a phase shift to the qubit of order $\Omega/\mu$, which is typically on the order of $0.1$ in the experiments reported here.  However, as will be discussed below this phase shift is greatly reduced by pulse shaping, and only enters quadratically into the total gate error.  The gate is operated by choosing $\mu$ near ($\sim 20$kHz) detuned from the axial stretch mode; ignoring all terms above that oscillate faster than $\delta$ leads to \eref{eq:simple_hamiltonian} in the main text.

\subsection{Numerical gate simulations}

Because we symmetrically detune the two lasers within the excited state manifold, the detuning $\Delta$ is controlled entirely by the magnetic field.  For the reported experiments, we chose a field of $5.57$Gauss (corresponding to a single-photon detuning $\Delta\approx 2\pi\times 7.8$MHz). For this detuning, achieving a two loop gate in $t_{\rm gate}\approx100\mu$s requires $|g|\approx2\pi\times 0.76$MHz; thus we expect excited state populations to be on the order of $(|g|/\Delta)^2\sim 1\%$, making perturbative adiabatic elimination not very well justified (at the level of precision we care about for high-fidelity gates).  Therefore, we chose to directly simulate the Hamiltonian as written in Eqs.\ (\ref{eq:fullham_init}-\ref{eq:fullham_fin}), with the excited-state structure fully intact.

The Hilbert space dimension is given by
\begin{align}
\mathcal{D}=\mathcal{D}_{\rm ion}\times\mathcal{D}_{\rm phonon}=4^2\times\prod_{\alpha=1}^{6}(n_{\rm max,\alpha}+1),
\end{align}
with $n_{\rm max,\alpha}$ the Fock-space occupation cutoff for mode $\alpha$.  Keeping $n_{\rm max}=10$ for the gate mode (which we find to be sufficient) and $n_{\rm max}=1$ for all other modes leads to $\mathcal{D}=5600$.  Considering the poor conditioning of the Schr\"odinger equation (large separation in time scales) caused by keeping the excited states, we found that numerical simulations in this full space are feasible but very time consuming. Instead, we assume that errors due to the spectator modes are small and uncorrelated to lowest order, justifying the following procedure: (1) calculate the gate error by including only the c.o.m.\ as a spectator mode and optimizing over the laser power, and then (2) repeat the calculation for each of the four radial spectator modes at the same parameters, each time calculating their population at the final gate time, (3) add those populations to the gate error estimate obtained in (1).  We note that the radial-sideband contributions to the gate error are relatively small compared to the contribution from the axial c.o.m., and in principle could be greatly reduced by moving the beams towards a counter-propagating geometry.


\subsection{Estimates of experimental error}

Leakage errors are quantified by preparing both ions in the $\ket{\uparrow}$ state and repeatedly applying the gate sequence, except that we apply only one gate beam or the other to quantify leakage due to each beam independently.  In addition to the leakage caused by spontaneous emission, we expect to accumulate some population in the $D$ state (or into other hyperfine states if it relaxes) due to failure of pulse shaping.  We observe a leakage rate of about $3\times 10^{-4}$ per gate, only about one quarter of which is explained by spontaneous scattering during the gate time.  Subtracting out leakage due to spontaneous emission, we find a leakage rate for the two different beams of $R_{A}\approx1.7\times 10^{-4}$ and $R_{B}\approx0.7\times 10^{-4}$.  The discrepancy between the two beams is consistent with an expected $1\times 10^{-4}$ leakage rate for beam $A$ due to a weak direct coupling from $\ket{\uparrow}\rightarrow\ket{e_0}$ (for beam $B$ this coupling is nominally forbidden by both the polarization and $k$-vector choices, and is extremely small), which we infer from a measured Rabi frequency for that transition of $2\pi\times 6\,$kHz. The remaining leakage of slightly more than $1\times 10^{-4}$ per gate is not understood, but may be the result of either (1) fast laser phase noise (on the timescale of $\Delta$) that interferes with the efficacy of pulse shaping and (2) heating during the long sequences used to measure this leakage, which increases coupling to sidebands of the excited states (finite mode coherence times will prevent all population from being pulse shaped out of these excited-state sidebands).

Laser phase noise enters into the gate infidelity in several possible ways depending on how fast it is.  Noise that is fast compared to the inverse single-photon detuning will reduce the efficacy of pulse shaping and lead to leakage, as described above.  Noise that is slow compared to this time scale will primarily impact the gate via induced fluctuations in the time-averaged AC-Stark shift.  We quantify this effect by performing a Ramsey spin-echo experiment on the clock transition $\ket{\uparrow}\rightarrow\ket{e_0}$ using the gate lasers (using a clock transition enables us to isolate the effect of laser phase noise from magnetic field noise) with the same duration as the gate time.  The probability to return to the incorrect state after the Ramsey sequence can be related to laser phase noise by the filter function $\mathcal{F}(\omega)$ for the spin-echo sequence as
\begin{align}
\epsilon_{\rm ramsey}=\int_{0}^{\infty}\mathcal{S}_{\phi}(\omega)\mathcal{F}(\omega)d\omega.
\end{align}
Here $\mathcal{S}_{\phi}(\omega)$ is the spectral density of phase fluctuations of the laser.  During the gate, laser frequency fluctuations $\delta\omega(t)$ that are slow compared to $1/\Delta$ result in a fluctuating AC-Stark shift, with the relation $\delta\Delta_{\rm LS}(t)\approx \delta\omega(t)\big[4 g^2/\Delta^2\big]$.  Since the gate is performed with a spin echo separating the same wait times as the Ramsey sequence described above, gate errors due to the AC-Stark shift fluctuations can be written in terms of the same filter function as above.  The only difference is that the spectral density of noise gets an overall scale factor of $\big[4 g^2/\Delta^2\big]^2$, which enables us to extract an expected gate error due to laser phase fluctuations of $\epsilon_{\phi}=\epsilon_{\rm ramsey}\times\big(4g^2/\Delta^2\big)^2\approx 3\times 10^{-4}$.

The effect of gate mode heating is calculated using the expression in Ref.\,\cite{PhysRevLett.117.060504} and an experimentally inferred upper bound on the gate mode heating rate.  Heating of the c.o.m.\ mode is also considered because this mode has a relatively large transient population during the gate (see \fref{fig:off_resonant_pops} in the main text), and the heating rate is large.  Assuming that absorption of a single phonon prevents pulse shaping from removing this population, we obtain an error estimate $\epsilon_{\rm c.o.m.}\sim \kappa_{\rm c.o.m.}t_{\rm gate}\times P_{\rm c.o.m.}$, where $\kappa_{\rm c.o.m.}$ is the heating rate of the c.o.m.\ mode and $P_{\rm c.o.m.}$ is its transient population.  We measure $\kappa_{\rm com}\approx 3\times 10^{3}$ quanta/s, and numerically calculate a transient population of $P_{\rm com}\approx 3\times 10^{-4}$, giving an error of about $1\times 10^{-4}$ for a $100\,\mu{\rm s}$ gate.

Errors due to imperfect microwave pulses are estimated by single-qubit randomized benchmarking of the microwave gates~\cite{Baldwin_2019}. We further verify the contribution of microwave gates to SRB by repeating the SRB sequences without LS gates and choosing an appropriate final gate, which consists only of microwaves, to ideally return to the initial state. We call this procedure SRB gap benchmarking and while it cannot be used to extract a fidelity of the microwave gates it is useful for interpreting the effects the single-qubit gates on the return probability in SRB. Under a simple single-qubit depolarizing error model, we can calculate the expected survival in SRB gap benchmarking from  single-qubit randomized benchmarking data. For 10 and 65 qutrit-Clifford gates this yields an expected survival of 0.982 and 0.965 compared to the measured return probability from SRB gap benchmarking of 0.984(4) and 0.955(9). Therefore, for SRB with LS gates we expect the single-qubit randomized benchmarking estimate to accurately estimate the effects of the microwave gates.

\subsection{Experimental fidelity estimation} 
  
The SRB procedure used in this paper is similar to the SRB-\textit{lite} procedure proposed in Ref.\,\cite{Baldwin_2019}. In SRB-\textit{lite}, leakage errors, which cause population to move outside of the symmetric subspace, are ignored. This is a good approximation for the current work since state preparation errors and leakage are believed to be small, and therefore do not contribute much to the fidelity. Even with leakage, the estimate from SRB-\textit{lite} provides an estimate of error rates across most of the symmetric subspace. The one difference between SRB-\textit{lite} and the procedure used here is that we fix the asymptote of the decay curve to $1/3$ when fitting instead of leaving it as a free parameter. This reduces the number of fit parameters and allows us to fit data with shorter sequence lengths. Leakage or SPAM errors may cause the asymptote to be less than $1/3$ but we believe these effects to be small. From numerical simulations we see fixing the asymptote to $1/3$ actually tends to return a lower fidelity estimate than leaving it as a free parameter. We ran this SRB procedure with three sequence lengths [1, 3, 7] each with 33 random sequences.

\end{document}